\documentclass[conference]{IEEEtran}
\IEEEoverridecommandlockouts
\usepackage{cite}
\usepackage{amsmath,amssymb,amsfonts}
\usepackage{algorithmic}
\usepackage{graphicx}
\usepackage{textcomp}
\usepackage{xcolor}
\def\BibTeX{{\rm B\kern-.05em{\sc i\kern-.025em b}\kern-.08em
    T\kern-.1667em\lower.7ex\hbox{E}\kern-.125emX}}
\newcommand{\tabincell}[2]{\begin{tabular}{@{}#1@{}}#2\end{tabular}}
\usepackage{graphicx}
\usepackage{subfigure}
\begin{document}

\title{A neural network approach to GOP-level rate control of x265 using Lookahead\\
}
\author{
    \IEEEauthorblockN{1\textsuperscript{st} Boya Cheng}
    \IEEEauthorblockA{\textit{School of Information and Communication Engineering} \\
    \textit{
        Communication University of China}\\
        Beijing, China \\
        boya\rule[-2pt]{0.15cm}{0.5pt}cheng@cuc.edu.cn}
\and
\IEEEauthorblockN{2\textsuperscript{nd} Yuan Zhang}
\IEEEauthorblockA{\textit{School of Data Science and Media Intelligence} \\
\textit{Communication University of China}\\
Beijing, China \\
yzhang@cuc.edu.cn}
}

\maketitle

\begin{abstract}
To optimize the perceived quality under a specific bitrate constraint, multi-pass encoding is usually performed with the rate control mode of the average bitrate (ABR) or the constant rate factor (CRF) to distribute bits as reasonably as possible in terms of perceived quality, leading to high computational complexity. In this paper, we propose to utilize the video information generated during the encoding to adaptively adjust the CRF setting at GOP level, ensuring the bits of frames in each GOP are allocated reasonably under the bitrate constraint with a single-pass encoding framework. In particular, due to the inherent relationship between CRF values and bitrates, we adopt a shallow neural network (NN) to map video content features to the CRF-bitrate model. The content-related features are collected from the lookahead module inside the x265 encoder, including encoding cost estimation, motion vector and so on. Further, a rate control method, called content adaptive rate factor (CARF), is proposed to adjust the CRF value of each GOP with the requirement of the target bitrate by using the predicted CRF-bitrate models of each GOP. The experimental results show that the proposed approach can make 84.5\% testing data within 20\% bitrate error (or better) and outperform the ABR mode in x265, leading to 5.23 \% BD-rate reduction on average.
\end{abstract}


\section{Introduction}
In rate control of x265, the average bitrate (ABR) mode ensures that the output stream achieves a predictable long-term average bitrate but it may fail in achieving the acceptable quality of each frame. In contrast, the constant rate factor (CRF) mode has the advantage of maintaining a certain level of perceived quality among frames by compressing different frames with different bits over the entire sequence. Its disadvantage is that the resulting bitrate is unpredictable\cite{b1}.

To ensure the optimal output quality under a certain bitrate constraint, multi-pass encoding method is usually used with both ABR and CRF mode. The former efficiently allocates the bits available based on the encoding cost statistics of previous passes. The latter adjusts the CRF value manually to make the resulting bitrate gradually approaches the target one. However, those method are time-consuming and mostly empirical.

The unpredictability of the ABR and CRF mode in terms of quality and bitrate is due to that the video content is changing over the entire sequence. Covell et al. discover a clear relationship existed among the bitrate, CRF value, frame resolution and frame rate. They develop a video content-related linear model to bridge between the bitrate and CRF value, and embed this model into a CRF prediction classification neural network to provide reliable guidance. The input of this network is content-related video features collected from previous transcoding results, such as the average number of bits used for the texture per macroblock (MB), motion vector (MV) per predicted MB and so on. To provide precise video information, at least twice transcoding are needed.

Sun et al. implement a regression neural network in a two-pass encoding framework to simplify the solution above. This network predicts content-dependent parameters of the CRF-bitrate model of each video segments instead of the CRF value directly. Then a CRF value is derived from the predicted model parameters with a given target bitrate. This solution improves the prediction accuracy by fixing the frame resolution and replacing the linear model with a second-order model. Besides that, the video feature collected from the first pass encoding result is similar to those used in the solution above.

All existing solutions are implemented outside the encoder and undertake sequence-level rate control. Furthermore, content-related features are gathered from previous transcoding or encoding statistic results so that those techniques have two pain points:
\begin{itemize}
\item First, the training data of the NN are 5s video segments sampled from long videos, which results in more than one scene may be included in one video segment. It means the determined CRF value is perhaps applied to different video contents.
\item Second, even if the above problem can be avoided, extracting video features from encoding results means that at least one pre-encoding or pre-transcoding is required, which is still multi-pass encoding essentially.
\end{itemize}

To optimize perceived quality under a certain bitrate constraint in a single-pass encoding framework, we attempt to adapt the GOP-level CRF setting according to the specified target bitrate. At the beginning of encoding, the encoder makes slice-type decision in batches using a module called lookahead. This module implements a scene-cut detection technique to determine whether the current frame is a scene switching based on the intra-frame and inter-frame encoding cost estimation about surrounding down-sampling frames.

On the one hand, the lookahead module divides a video into different GOPs, mainly depending on the encoder’s flexibility to the scene change. It is worth noticing that lots of hyperparameters influence the structure of GOP and the position of scene-cut so that the related hyperparameters jointly determine the flexibility of encoders before encoding. Overall, adjusting the CRF setting at the GOP level can achieve a finer granularity of adaptive selection.

On the other hand, in order to estimate the encoding cost, the lookahead module performs intra coding and inter coding on subsampled frames respectively with fixed block size including fast motion search and estimation. The analysis result generated during this stage plays a vital role in other encoding tools. For example, the reference picture list information is used by Macro Block-tree (MB/CU-Tree) technique\cite{b4} to calculate the block-level quantization parameter (QP) offset, the ABR and CRF mode leverage the encoding cost to derive the frame-level initial QP. However, none of the frames-level rate control technique has benefited from lookahead yet.

Accordingly, in this paper we propose a content adaptive rate factor (CARF) rate control approach which attempts to take advantage of the video content information of GOPs,  generated in lookahead, to decide the CRF value for each corresponding GOP. Only one encoding is required to enable the GOP level CRF setting to be controlled by the target bitrate directly.

The rest of this paper is organized as follows. Section II introduces our proposed framework as well as the key points in this solution. Section III evaluates the proposed method in terms of prediction accuracy and encoding performance. Finally, discussion and conclusion are drawn in Section IV.

\section{The Content Adaptive Factor Scheme}
\subsection{The proposed framework}\label{AA}
\begin{figure}[tb]
    \centerline{\includegraphics[width=\linewidth]{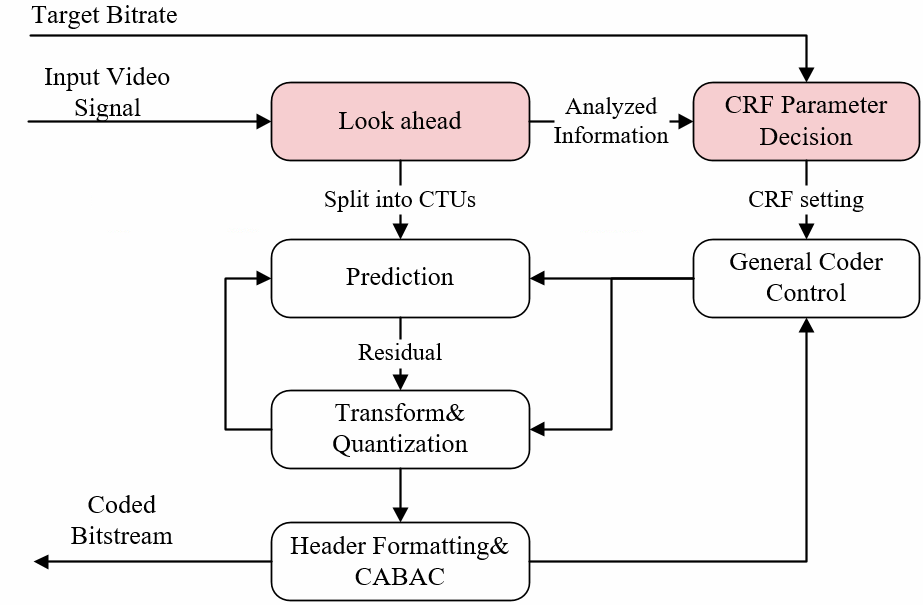}}
    \caption{Overview of the proposed framework}
    \label{fig1}
\end{figure}

Fig.~\ref{fig1}illustrates the proposed single-pass encoding framework . When the current frame is an I/IDR frame, CRF Parameter Decision module outputs a CRF value based on the current lookahead analysis result and a given target bitrate. Then, this CRF value is fed into General Coder Control module to calculate the frame level base QP of each frame in current GOP. The current CRF value remains the same until next I/IDR frame is decided.
\subsection{Content-dependent features collected from Lookahead}
Except for the coding and perceptual redundancy, the main task of the hybrid encoding framework is to eliminate the spatial-temporal redundancy. The spatial-temporal complexity of videos could influence the encoding parameters configuration and the encoding performance greatly. Hence we choose the video features mostly from the lookahead information representing spatial-temporal characteristics. The final features are described as follows:
\begin{itemize}
\item a score of prediction encoding cost
\item sum of pixels for each channel (Y, U, V)
\item square sum of pixels for each channel
\item a score of AC per MB
\item percentage of I MBs in P frame
\item the length of the MV for each predicted MB
\end{itemize}
We also take the original bitrate and frame rate into consideration while assuming that the frame size is fixed since the experiment results of\cite{b2}\cite{b3} have suggested that fixing resolution in prediction task is a reasonable choice.

\subsection{The CRF-Bitrate model in the CRF Parameter Decision module}
The CRF Parameter Decision module is driven by a regression prediction neural network (NN), which is used to predict parameters of the CRF-bitrate model. With a target bitrate, this module can compute a CRF value based on the predicted CRF-bitrate model directly.

After multiple fitting tests on hundreds of video clips, we confirm that the second-order model between CRF and bitrate is more accurate than the linear model as Sun et al. proposed. Therefore, we employ a content-dependent equation to model the relationship between CRF and bitrate. Given a bitrate $R$, the corresponding CRF setting of GOP  $g$ in the video $v$ is as follows:
\begin{equation}
crf(v,g)= a(v,g)ln(R)^{2} + b(v,g)ln(R) + c(v,g)\label{eq1}
\end{equation}
where $a(v,g)$, $b(v,g)$, and $c(v,g)$ are the content-dependent parameters which are the predication target of the regression network.

\subsection{The neural network in the CRF Parameter Decision module}
We adopt a shallow fully connected neural network with two hidden layers to predict the parameters of the CRF-Bitrate model by feeding scaled video features.
\begin{figure}[tb]
    \centerline{\includegraphics[width=\linewidth]{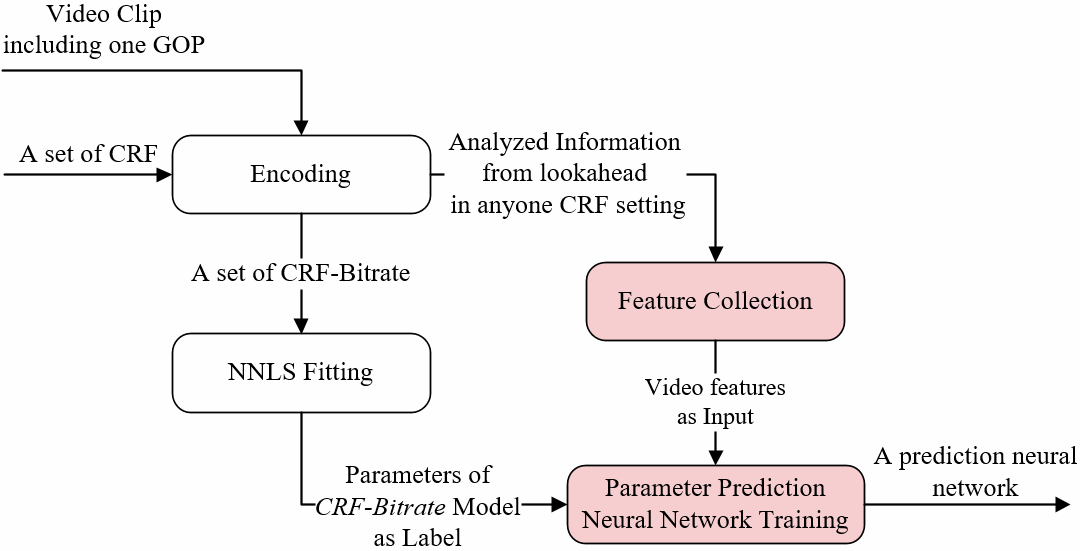}}
    \caption{Layout of neural network used to predict parameters of CRF-Bitrate model.}
    \label{fig2}
\end{figure}

As shown in Fig.~\ref{fig2}, with a set of target bitrate (from 0.2Mbps to 12Mbps), a set of CRF values $\hat{crf}_i$ are computed according to the predicted parameters and a set of CRF values $crf_i$ are computed by the parameter labels. The mean absolute error of those CRF setting is the final loss as the \eqref{eq2} shown.
\begin{equation}
  loss = \frac{1}{n} \sum_{i=0}^{n} |crf_i - \hat{crf_i}|\label{eq2}
\end{equation}

\begin{figure}
  \centering
  \includegraphics[width=\linewidth]{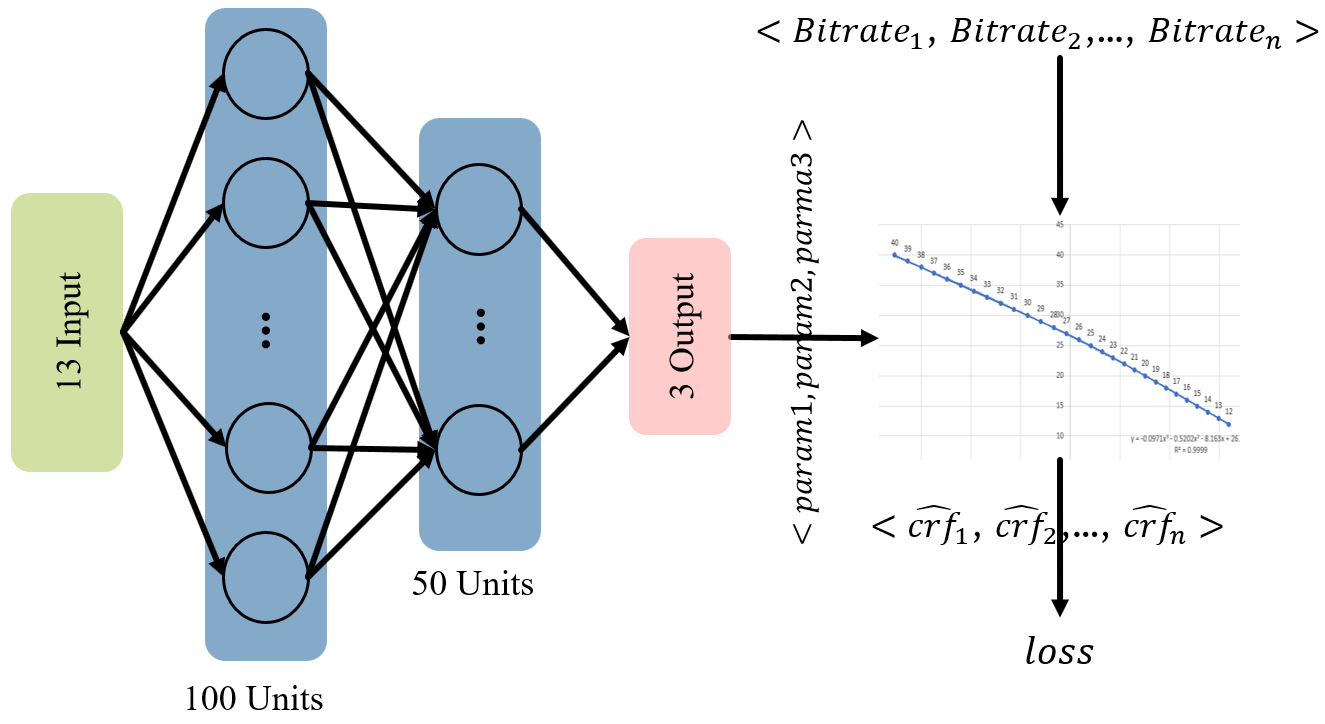}
  \caption{The training procedure of the regression prediction neural network}
  \label{fig3}
\end{figure}

To train the NN, we take the parameters of CRF-Bitrate models fitted by non-negative least squares (NNLS) as ground truth, which is fitted with 15 different CRF settings(from 12 to 40, with an interval of 2). Fig.~\ref{fig3} depicts the block diagram of the training procedure of the proposed NN.
\section{Experiments}
In this section, we first evaluate our method in regression task with a set of target bitrates (0.3, 0.75, 1.2, 1.85, 2.85, 4.3Mbps), to demonstrate the benefit of using the lookahead information to represent the video characteristics and decide the CRF setting. Then we evaluate the encoding performance of the proposed CARF mode in encoding task with a set of target bitrates (0.5, 0.75, 1.5 and 3.5 Mbps).
\subsection{Dataset}
To train the NN, we collect the User-generated content (UGC) videos with fixed resolution (720p, including 1280x720 and 720x1280) from a commercial video-sharing website\footnote{https://www.bilibili.com/}. Those videos mainly belong to the video blog (vlog) which has no specific theme, enabling the dataset content-rich. We also consider some typical content scenarios such as fast-moving and dark scene.

Since CRF values are configured at GOP-level, no scene-cut should exist in the video clips of dataset. In random access (RA) coding structure, we sample short video clips from collected videos by FFmpeg\cite{b5} to keep the frame numbers of video clips ranging from 75 to 140 that corresponds to the configuration of rc-lookahead. Furthermore, we manually filter out the clips with more than one scenes. In the end, 5031 video clips are sampled from 600 videos. 80\% video clips are used for training and 20\% are for validation.

\subsection{Hyperparameter}
The whole experiment has been tested on the PC with Intel$^{\circledR}$ Core$^{TM}$  i7-6850k 3.60GHz processor and 62.0GB system memory. The proposed method is implemented in x265 version2.9\cite{b6}. The real encoding is performed under the following configurations:
\begin{itemize}
\item preset: medium, which is the default setting
\item tune: psnr, which disables the perceptual optimization
\item rc-lookahead: 100, which is the number of frames for slice-type decision lookahead
\item min-keyint: 40, which is the minimum GOP size
\end{itemize}

The encoding has been carried on single threading by removing all parallel encoding default configurations including frame-level parallel and wave-front parallel processing (WPP).
\subsection{ Accuracy result in terms of regression prediction}
The accuracy of the proposed nonlinear regression network is evaluated by the bitrate error as follows:
\begin{equation}
    bitrate_{error} = \frac{|R_a - R_t|}{R_t} \ast 100\%
    \label{eq3}
\end{equation}
Where $R_a$ is the actual resulting bitrate, $R_t$ is the target bitrate.

\begin{figure}[tb]
    \centerline{\includegraphics[width=\linewidth]{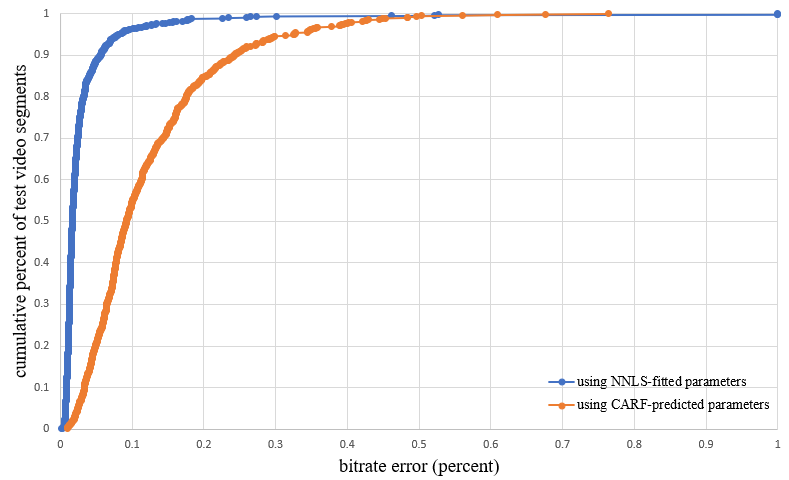}}
    \caption{Cumulative distribution of bitrate errors(\%) across the testing set}
    \label{fig4}
\end{figure}

Fig.~\ref{fig4} illustrates the cumulative distribution of the bitrate error of NNLS-fitted labels (blue curve) and the proposed neural network (orange curve), respectively. The result of NNLS-fitted labels can be seen as the upper bound of our approach. Our approach keeps 84.5\% of our testing data at or below the 20\% bitrate error, as shown in Fig.~\ref{fig4}.

Table.~\ref{tab1} shows the comparison between our approach and the reference methods\cite{b2}\cite{b3} in terms of prediction accuracy, decision granularity and requirements of methods. In terms of prediction accuracy, our method outperforms the approach proposed by YouTube\cite{b2} but is inferior to UMKC\cite{b3}. However, it is worth noticing that Sun et al. employ a specific CRF setting on each resolution to obtain labels for NN training. If this CRF setting is close to the setting of target bitrates for testing, prediction accuracy will become higher. It means that the selection of the CRF setting during the ground truth generation phase may influence the final prediction result. Therefore, the instability of Sun’s method will reduce its performance.
\begin{table}[tbhp]
    \caption{The Prediction Performance}
    \begin{center}
    \begin{tabular}{|c|c|c|c|}
    \hline
    & \textbf{\textbf{YouTube[2]}} & \textbf{UMKC[3]} & \textbf{CARF}\\
    \hline
    \tabincell{c}{\textbf{Error}\\ \textbf{within 20\%}} & 80\%& 91.6\%&  84.5\%\\
    \hline
    \tabincell{c}{\textbf{Decision}\\ \textbf{Granularity}}&Segment level&Segment level&GOP level\\
    \hline
    \textbf{Requirement}& \tabincell{c}{Two pre\\-transcoding\\are needed} & \tabincell{c}{One pre\\-transcoding\\is needed}&  No need\\
    \hline
    \end{tabular}
    \label{tab1}
    \end{center}
\end{table}

Considering all-of-the-above factors listed in Table I, it can be seen that the lookahead information represents the spatiotemporal properties of videos well and the proposed one-pass approach is feasible and effective.
\subsection{Encoding performance evaluation}
To evaluate the encoding performance, we compute the BD-rate\cite{b7} of the proposed rate control algorithm and take the ABR mode as the anchor, because both of them accept specified bitrate as the input for rate control. We evaluate on 12 samples of 6-7s video segments, which belong to the same platform where the dataset is collected. Table.~\ref{tab2} lists the description of those segments.
\begin{table}[t]
    \caption{Description of Testing Videos}
    \begin{center}
        \begin{tabular}{|c|c|c|c|}
        \hline	
        \tabincell{c}{\textbf{Sequence}\\ \textbf{Number}}& \textbf{Inside$\backslash$Outside} & \textbf{Bright$\backslash$Dark} & \textbf{moving speed}\\
        \hline
        01&inside&bright& medium\\
        \hline
        02&outside&bright& medium\\
        \hline
        03&inside&bright& slow\\
        \hline
        04&outside&dark& fast\\
        \hline
        05&outside&bright& medium\\
        \hline
        06&outside&dark& fast\\
        \hline
        07&outside&dark& medium\\
        \hline
        08&inside&bright& medium\\
        \hline
        09&inside&bright& veryfast\\
        \hline
        10&inside&bright& veryfast\\
        \hline
        11&concert&bright& fast\\
        \hline
        12&concert&dark& medium\\
        \hline
        \end{tabular}
        \label{tab2}
    \end{center}
\end{table}

The respective testing results compared with single-pass and two-pass ABR are shown in Table.~\ref{tab3}. These results demonstrate that the proposed CARF mode performs better than the ABR mode, leading to average 4.12\% BD-rate reduction in PSNR, 5.35\% BD-rate reduction in VMAF and 5.73\% BD-rate reduction in SSIM (over single-pass ABR).
\begin{table}[t]
    \caption{Result of BD-Rate(\%) over x265-ABR}
    \begin{center}
        \begin{tabular}{|c|c|c|c|c|c|c|}
            \hline
            & \multicolumn{6}{|c|}{\textbf{Random Access Main 10}}\\
            \hline
            \textbf{Reference}& \multicolumn{3}{|c|}{\textbf{Over x265-single-passABR}}&
            \multicolumn{3}{|c|}{\textbf{Over x265-two-passABR}}\\
            \hline
            \tabincell{c}{\textit{Sequence}\\ \textit{Number}}&\textit{PSNR}&\textit{VMAF}&\textit{SSIM}&
            \textit{PSNR}&\textit{VMAF}&\textit{SSIM}\\
            \hline
            01&	-5.25&	-11.94&	-8.13&	-9.21&	-9.40&	-4.82\\
            \hline
            02&	-3.00&	-6.51&	-5.23&	-1.63&	-2.50&	0.48\\
            \hline
            03&	-4.02&	-7.81&	-3.99&	-2.70&	-2.05&	-0.40\\
            \hline
            04&	-1.82&	-5.17&	-3.12&	-4.07&	-0.96&	-2.53\\
            \hline
            05&	-1.06&	-2.40&	-1.21&	-1.55&	-1.63&	1.30\\
            \hline
            06&	-7.09&	-16.39&	-11.11&	-10.68&	-11.25&	0.96\\
            \hline
            07&	-4.29&	-6.57&	-6.06&	-5.83&	-6.11&	1.17\\
            \hline
            08&	-1.87&	-5.53&	-4.72&	-4.60&	-2.33&	-2.43\\
            \hline
            09&	-15.02&	13.38&	-12.02&	-3.85&	-0.14&	2.75\\
            \hline
            10&	-1.73&	-5.02&	-2.43&	-4.59&	-7.50&	-4.60\\
            \hline
            11&	-1.29&	-2.36&	-3.61&	-0.15&	-0.76&	-1.48\\
            \hline
            12&	-2.96&	-7.91&	-7.13&	-3.96&	-0.79&	-0.95\\
            \hline
            \textbf{Overall}&	-4.12&	-5.35&	-5.73&	-4.40&	-3.78&	-0.88\\
            \hline
            \end{tabular}
        \label{tab3}
    \end{center}
\end{table}

\begin{figure}[htbp]
\centering
\subfigure[Bits fluctuation: 2-pass ABR $>$ CARF $>$ CRF]{
\includegraphics[width=\linewidth]{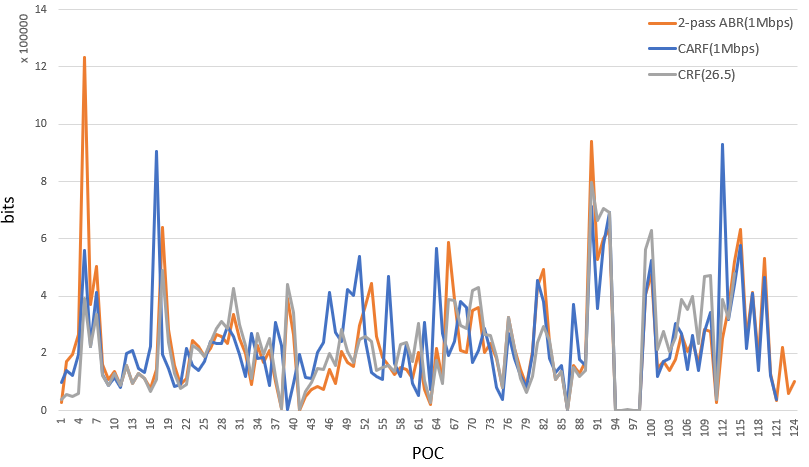}
}
\quad
\subfigure[Quality fluctuation: 2-pass ABR $>$ CARF $>$ CRF]{
\includegraphics[width=\linewidth]{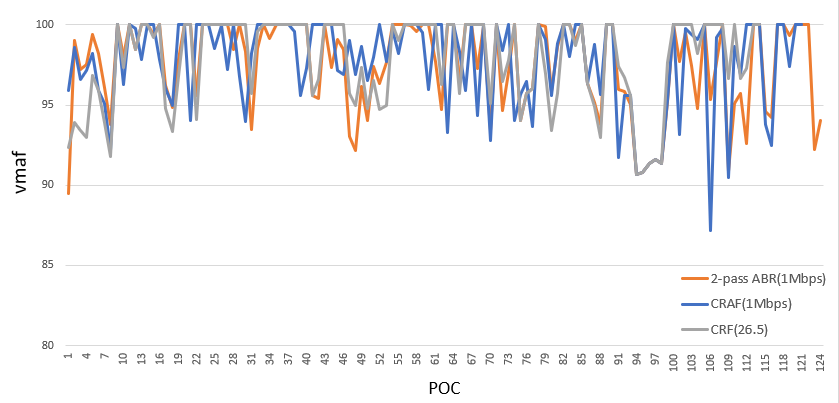}

}
\caption{The fluctuation of the bits and quality over a 10mins video sequence at 1Mbps (2-pass ABR:orange;  CARF:blue; CRF:gray)}
\label{fig5}
\end{figure}

Given a target bitrate, it is so empirical to find the suitable CRF value for each testing video that the BD-rate over CRF is not be computed. But we compare the CRF, 2-pass ABR with CARF in terms of the fluctuation of bit number and quality over several long videos. The CRF outperforms others and CARF is better than ABR. But the selection of CRF value is a tricky problem  whereas the CARF can balance the BD-rate and target bitrate well. Fig.~\ref{fig5} shows the result over one of the long video.
\subsection{Encoding time evaluation}
We also evaluate the encoding time as follows. For each segment $v$,  its running time $Ent_v$ is the geometric mean of encoding time percent $Ent_{v,r}$, which is derived as \eqref{eq4}.
\begin{equation}
    Enr_{v,r} = \frac{|R_{carf,r} - R_{abr,r}|}{R_{abr,r}} \ast 100\%
    \label{eq4}
\end{equation}
where the $R_{carf,r}$ and $R_{abr,r}$ is the encoding time consumed by CARF and ABR with a specific target bitrate $r$.

Table.~\ref{tab5} shows that the running time of CARF decreases by 31\% over the two-pass ABR but increases by 48\% over the single-pass ABR on average. Our current implementation can be further optimized to reduce the computational complexity.
\begin{table}[tbhp]
    \caption{Result of Encoding Runtime(\%) over x265-ABR}
    \begin{center}
        \begin{tabular}{|c|c|c|}
        \hline
        & \multicolumn{2}{|c|}{\textbf{Random Access Main 10}}\\
        \hline
        \tabincell{c}{\textit{Sequence}\\ \textit{Number}}&\tabincell{c}{\textit{Over x265-}\\ \textit{single-pass-ABR}}&
        \tabincell{c}{\textit{Over x265-}\\ \textit{two-pass-ABR}}\\
        \hline
        01&	125&	66\\
        \hline
        02&	144&	77\\
        \hline
        03&	120&	65\\
        \hline
        04&	126&	73\\
        \hline
        05&	154&	81\\
        \hline
        06&	146&	82\\
        \hline
        07&	125&	68\\
        \hline
        08&	136&	71\\
        \hline
        09&	230&	119\\
        \hline
        10&	216&	115\\
        \hline
        11&	124&	65\\
        \hline
        12&	133&	71\\
		\hline
        \textbf{Overall}&	148&	79\\
        \hline
        \end{tabular}
        \label{tab5}
    \end{center}
\end{table}
\section{Conculsions}

In this paper, we have proposed a GOP-level rate control scheme called CARF to explore how to relate the target bitrate with CRF directly in one-pass encoding framework. This scheme utilizes the lookahead analysis result inside the encoder to configure suitable CRF parameter setting for each GOP under target bitrate. The experimental results have shown that the CARF can keep 20\% bitrate error (or lower) on 84.5\% testing data. Compared with the ABR of x265, the CARF has 4.12\%, 5.35\%, 5.73\% BD-rate reduction in terms of PSNR, VMAF and SSIM respectively(comparing with single-pass ABR in x265). In particular, this method has following advantages:
\begin{itemize}
    \item As shown in the experimental results, the lookahead information adopted in the proposed scheme can contribute greatly in CRF setting selection.
    \item By using the analysis results in the middle of encoding, we implement a single-pass encoding framework, which shows the rationality and effectiveness of the one-pass solution implemented inside codec.
    \item This approach changes the CRF parameter setting as the scene-cut is detected, enabling a finer CRF application granularity without extra effort.
\end{itemize}

According to some analysis from the experiments, we further explore below two implications for future work.
\begin{itemize}
    \item The relationship between CRF and the video quality can be obtained in the same way, which has been observed in our subsequent experiment with VMAF, PSNR and SSIM. With the CRF-quality and CRF-bitrate relationship being clear, a new rate control scheme can be explored by the same way under the joint constraint of bitrate and quality which is expected to benefit the practical scenarios more.
    \item As shown in Fig.~\ref{fig4}, there is still room to further improve the prediction accuracy of the neural network model, if considering the information in addition to the lookahead module. We can exploit the convolutional neural network (CNN) which is good at feature extraction so that wider and more features can be obtained to describe video characteristic comprehensively, including not only the spatiotemporal redundancy but also the visual one.
\end{itemize}
\section*{Acknowledgment}

This work was supported in part by the National Natural Science Foundation of China (Grant No.61971382) and the Fundamental Research Funds for the Central Universities (Grant No.CUC19ZD004)

\end{document}